# From first-order magneto-elastic to magneto-structural transition in (Mn,Fe)$_{1.95}$P$_{0.50}$Si$_{0.50}$ compounds

N.H. Dung,[a] L. Zhang, Z.Q. Ou, E. Brück

Fundamental Aspects of Materials and Energy (FAME), Faculty of Applied Sciences, Delft University of Technology, Mekelweg 15, 2629 JB Delft, The Netherlands

[a)] E-mail: h.d.nguyen@tudelft.nl

We report on structural, magnetic and magnetocaloric properties of Mn$_x$Fe$_{1.95-x}$P$_{0.50}$Si$_{0.50}$ ($x \geq 1.10$) compounds. With increasing the Mn:Fe ratio, a first-order magneto-elastic transition gradually changes into a first-order magneto-structural transition via a second-order magnetic transition. The study also shows that thermal hysteresis can be tuned by varying the Mn:Fe ratio. Small thermal hysteresis (less than 1 K) can be obtained while maintaining a giant magnetocaloric effect. This achievement paves the way for real refrigeration applications using magnetic refrigerants.

Nowadays, advanced magnetocaloric materials often undergo a first-order magnetic transition (FOMT),[1-4] because the FOMT is associated with an abrupt change in crystal lattice which enhances magnetocaloric effects (MCE) via a spin - lattice coupling. The FOMT can be divided into first-order magneto-structural transition (FOMST) which exhibits a structure change coupled with a magnetic transition as observed for Gd$_5$(Ge$_x$Si$_{1-x}$)$_4$,[5,6] Ni$_{0.50}$Mn$_{0.50-x}$Sn$_x$ (Ref. 7) and MnCoGeB$_x$;[8] or first-order magneto-elastic transition (FOMET) for which the crystal structure remains unchanged but the lattice constants suddenly change at the magnetic transition, as observed for MnFeP$_{1-x}$As$_x$ (Ref. 9) and La(Fe$_{1-x}$Si$_x$)$_{13}$.[10,11]

Fe$_2$P-based compounds are known as giant magnetocaloric materials with a FOMET. Most studies have recently focused on (Mn,Fe)$_2$(P,As,Ge) compounds.[1-3,9,12,13] However, the limited availability of Ge and toxicity of As hold these materials back from real refrigeration applications. Substitution of Si for As and Ge becomes one of the most prominent studies towards making a high performance room-temperature magnetic refrigerant. Some efforts have been made to reduce thermal hysteresis ($\Delta T_{hys}$) which was found to be very large ($\Delta T_{hys}$ = 35 K) in the MnFeP$_{0.50}$Si$_{0.50}$ alloy.[14] Here we report on (Mn,Fe)$_{1.95}$P$_{0.50}$Si$_{0.50}$ compounds when changing the Mn:Fe ratio with emphasis on the behavior of magnetic and structural transitions. We observe a previously unknown FOMST and a modified FOMET favorable for real refrigeration applications.

The (Mn,Fe)$_{1.95}$P$_{0.50}$Si$_{0.50}$ alloys were prepared by ball-milling. Proper amounts of Mn (99.9%), Si (99.999%) chips, binary Fe$_2$P (99.5%) and red-P (99.7%) powder were mixed and ball-milled for 10 hours. The fine powder was then pressed into small tablets and sealed in quartz ampoules in an Ar atmosphere of 200 mbar. The samples were sintered at 1373 K for 2 hours and then annealed at 1123 K for 20 hours before being oven cooled to room temperature. Magnetic measurements were carried out using the RSO mode in a SQUID magnetometer (Quantum Design MPMS 5XL). X-ray diffraction patterns were obtained by a PANalytical X-pert Pro diffractometer with Cu Kα radiation, secondary flat crystal monochromator and X'celerator RTMS Detector system.

The room-temperature X-ray diffraction measurements pointed out that all the samples crystallize in the hexagonal Fe$_2$P-type structure (space group P-62m). A very small amount of cubic (Mn,Fe)$_3$Si (space group Fm3m) was also detected and hardly affects the results of magnetic measurements.[15] The temperature dependence of the magnetization for the Mn$_x$Fe$_{1.95-x}$P$_{0.50}$Si$_{0.50}$ compounds measured in a field of 1 T is shown in Fig. 1a. For $x$ < 1.40, the M-T curves show very sharp ferro – paramagnetic transitions. A clear $\Delta T_{hys}$ confirms the first-order nature of these transitions. The $\Delta T_{hys}$ can be tuned from 5 K, 2 K to 1 K by varying the Mn:Fe ratio from $x$ = 1.20, 1.25 to 1.30, respectively. Figure 1b shows the isothermal magnetic entropy change ($\Delta S_m$) as a function of temperature under a field change $\Delta B$ of 0-1 T and 0-2 T for the $x$ = 1.20, 1.25, 1.30, 1.40 samples. Here the $\Delta S_m$ is calculated using magnetization isotherms through the Maxwell relations.[4] Apparently, the absolute value of $\Delta S_m$ is lower in the sample with more Mn. However, it should be noted that the $x = 1.30$ sample which has a very small $\Delta T_{hys}$ still displays a large $|\Delta S_m|$ of 15 Jkg$^{-1}$K$^{-1}$ under a 2 T field change. This value is 4 times higher than that of the benchmark material Gd.[16]

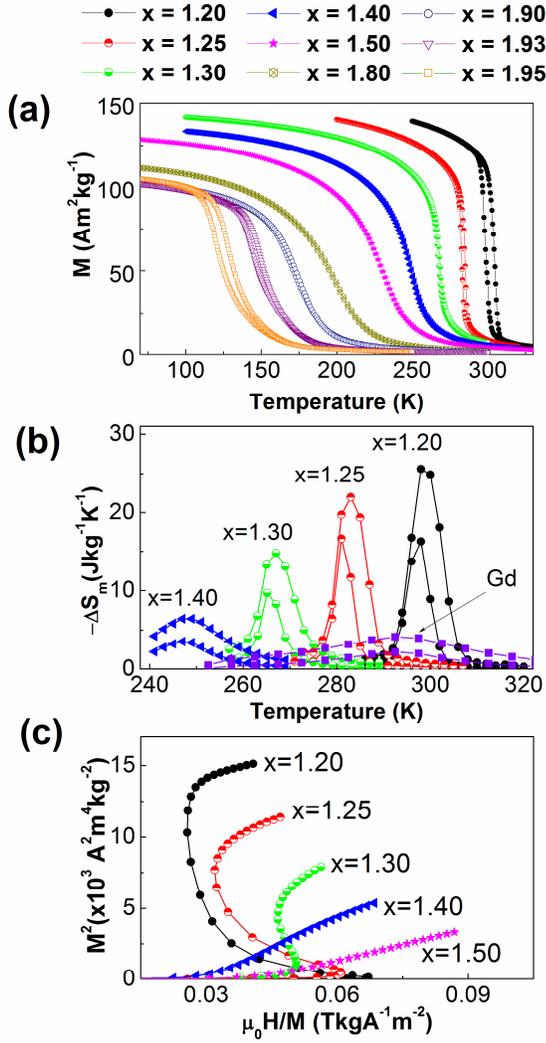

Figure 1. (Color online) Temperature dependence of the magnetization in a field of 1 T on cooling and heating (a), the magnetic entropy change as a function of temperature under a field change of 0-1 (lower curves) and 0-2 T (upper curves) (b), and Arrot plots obtained from increasing field magnetization isotherms in the vicinity of the transition temperature (c) for the $Mn_xFe_{1.95-x}P_{0.50}Si_{0.50}$ compounds. The magnetic entropy change of the benchmark material Gd is added for comparison.

Figure 1c illustrates Arrot plots derived from the magnetization isotherms in the vicinity of the transition temperature for the $x = 1.20, 1.25, 1.30, 1.40$ and $1.50$ samples. The S-shaped magnetization curves revealing relevant high-order terms in the Landau free energy expansion[13] prove a FOMT for $x < 1.40$.

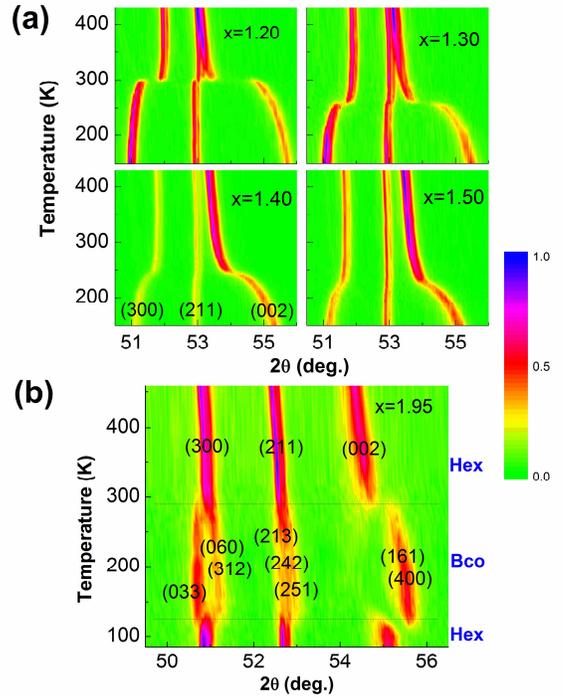

Figure 2. (Color online) The contour plots of the X-ray diffraction patterns for the $Mn_xFe_{1.95-x}P_{0.50}Si_{0.50}$ compounds: $x = 1.20, 1.30, 1.40, 1.50$ (a) and $x = 1.95$ (b) collected using Cu Kα radiation in a zero magnetic field on heating. Only the small range of $2\theta$ is shown for clarity. The bar on the right represents the normalized intensity scale.

However, neither a negative slope nor an inflection point is observed for the $x = 1.40$ and $1.50$ samples, confirming a second-order magnetic transition (SOMT). Thus, replacing some Fe with Mn can lower the energy barrier in the FOMT, and the FOMT gradually changes into a SOMT when the energy barrier becomes lower and finally vanishes.

X-ray diffraction at various temperatures confirms a stable hexagonal Fe$_2$P-type structure for $x = 1.20, 1.25, 1.30, 1.40, 1.50$ and $1.80$ samples. The thermal evolution of the X-ray diffraction patterns is illustrated in Fig. 2a for $x = 1.20, 1.30, 1.40$ and $1.50$. For the $x = 1.20$ and $1.30$ samples, a discontinuity of the diffraction peaks at the transition temperature indicates a jump of lattice constants. The sample with larger MCE exhibits a stronger peak shift at the critical temperature. Additionally, the (300) and (002) peaks are shifted in the opposite direction which implies that the lattice constants $a$ and $c$ change in opposite sense. The steep change of lattice constants coupled with the magnetic transition confirms a FOMET. On the other hand, a continuity of the peaks with respect to temperature is observed for the $x = 1.40$ and $1.50$ samples, demon-strating a SOMT. This observation is in good agreement with the results from the Arrot plots.

The ferromagnetic order of the $Mn_xFe_{1.95-x}P_{0.50}Si_{0.50}$ compounds remains until Fe is completely replaced by Mn (see Fig. 1a). Note that the existence of the $\Delta T_{hys}$ for $x \geq 1.90$ indicates the reoccurrence of a FOMT. However structural measurements show that these samples have a different type of FOMT. On cooling below room temperature we find that the paramagnetic hexagonal phase is transformed into a paramagnetic body-centered orthorhombic (bco) phase (space group $Imm2$). Further cooling makes this orthorhombic structure transform back into the hexagonal structure in a FOMST. The orthorhombic structure is similar to that found in $Fe_2(P,Si)$ compounds.[17,18] The re-entrant hexagonal structure in the FOMST points to a preference of the ferromagnetism for the hexagonal rather than the orthorhombic structure. Figure 2b illustrates the structural transformation on heating for the $x = 1.95$ sample without Fe. The orthorhombic phase exists over a temperature range from 125 K to 290 K between two ferromagnetic and paramagnetic hexagonal phases. A quite large structural transition zone at which both the hexagonal and orthorhombic phases exist explains why we could not see a sharp magnetic transition in the FOMST (see Fig. 1a). The MCE at the FOMST is therefore not so large with $|\Delta S_m| \sim 6$ $Jkg^{-1}K^{-1}$ under $\Delta B = 0\text{-}2$ T for the $Mn_{1.95}P_{0.50}Si_{0.50}$ compound.[15]

Since hexagonal $Mn_xFe_{2-x}P$ ($x > 1.35$) is antiferromagnetic,[19] the present study indicates that the Si addition supports a ferromagnetic order. The enhancement of the ferromagnetism by replacing P with Si was also observed in $Fe_2(P,Si)$.[17,18]

Because of a strong coupling between the crystal and magnetic structure, the magnetic-ordering temperature ($T_C$) can be defined as the temperature at which we see a very clear change in tendency of the thermal evolution of the lattice constants or the structure. In some cases of the FOMT, $T_C$ can also be determined from extremes of the first derivative of the isofield magnetization with respect to temperature. Shown in Fig. 3 is the phase diagram of the $Mn_xFe_{1.95-x}P_{0.50}Si_{0.50}$ compounds. It is found that the temperature range over which the orthorhombic phase exists becomes narrower and disappears as $x$ is lower than a critical value which is extrapolated to be about $x = 1.88$. $T_C$ decreases almost linearly with the Mn:Fe ratio for the FOMET, gradually for the SOMT and drops faster in a very narrow composition range of the FOMST. The Mn(3$g$)-Fe(3$f$) inter-layer exchange coupling is believed to be responsible for a ferromagnetic order in MnFe(P,Si) compounds.[20] For the Mn-rich compounds, since the Mn atom favors the 3$g$ site, excess Mn enters the 3$f$ site. The reduction of $T_C$ demonstrates that the replacement of Mn for Fe weakens the Mn(3$g$)-Fe/Mn(3$f$) inter-layer exchange coupling in the hexagonal Mn-rich compounds. If Mn is replaced by Fe at the 3$g$ site, the Mn/Fe(3$g$)-Fe(3$f$) exchange coupling will be stronger than that of the Mn-rich compounds. Therefore, the ferromagnetism will be enhanced and $T_C$ of hexagonal $Fe_2P_{0.8}Si_{0.2}$ with even less Si content (about 510 K) was found to exceed that of the Mn-rich compounds.[17,18]

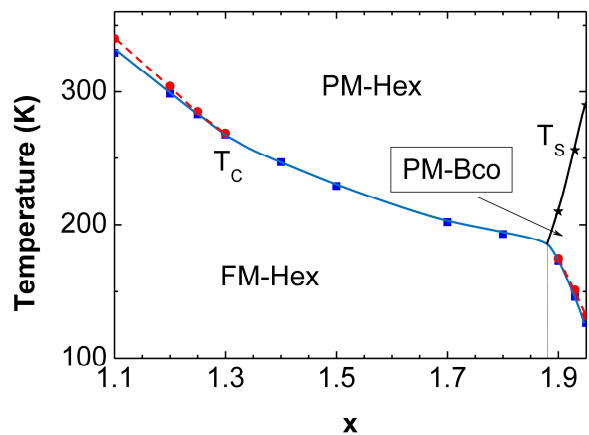

Figure 3. (Color online) Phase diagram of the $Mn_xFe_{1.95-x}P_{0.50}Si_{0.50}$ compounds. The red dashed and blue solid curves represent the $T_C$ on heating and cooling, respectively, the $T_S$ represents the paramagnetic orthorhombic – paramagnetic hexagonal transition temperature on heating.

In summary, by varying the Mn:Fe ratio the $(Mn,Fe)_{1.95}P_{0.50}Si_{0.50}$ compounds change from a FOMET into a FOMST through a SOMT. The study also shows the preference of the hexagonal $Fe_2P$-type structure for the ferromagnetic order and a strong coupling between the crystal and magnetic structure. Although the MCE from the FOMST is not so large, the existence of the FOMST with the re-entrant hexagonal structure is very interesting for further study on exchange coupling in $Fe_2P$-based materials. Moreover, small $\Delta T_{hys}$ with a giant MCE can be achieved in a FOMET favorable for real magnetic refrigeration applications in the near future.

### Acknowledgements

This work is part of the Industrial Partnership Program, Foundation for Fundamental Research on Matter (FOM), the Netherlands and co-financed by BASF Future Business.

### References

[1] E. Brück, J. Phys. D. Appl. Phys. **38**, R381 (2005).
[2] E. Brück, O. Tegus, D. T. C. Thanh, N. T. Trung, and K. H. J. Buschow, Int. J. Refrig. **31**, 763 (2008).

[3] E. Brück, O. Tegus, D. T. C. Thanh, and K. H. J. Buschow, J. Magn. Magn. Mater. **310**, 2793 (2007).

[4] K. A. Gschneidner, V. K. Pecharsky, and A. O. Tsokol, Rep. Prog. Phys. **68**, 1479 (2005).

[5] V. K. Pecharsky and K. A. Gschneidner, Phys. Rev. Lett. **78**, 4494 (1997).

[6] F. Casanova, A. Labarta, X. Batlle, F. J. Perez-Reche, E. Vives, L. Manosa, and A. Planes, Appl. Phys. Lett. **86,** 262504 (2005).

[7] T. Krenke, E. Duman, M. Acet, E. F. Wassermann, X. Moya, L. Manosa, and A. Planes, Nat. Mater. **4**, 450 (2005).

[8] N. T. Trung, L. Zhang, L. Caron, K. H. J. Buschow, and E. Brück, Appl. Phys. Lett. **96**, 162507 (2010).

[9] O. Tegus, E. Brück, K. H. J. Buschow, and F. R. de Boer, Nature (London) **415**, 150 (2002).

[10] S. Fujieda, A. Fujita, and K. Fukamichi, Appl. Phys. Lett. **81**, 1276 (2002).

[11] A. Fujita, S. Fujieda, Y. Hasegawa, and K. Fukamichi, Phys. Rev. B **67**, 104416 (2003).

[12] O. Gutfleisch, M. A. Willard, E. Brück, C. H. Chen, S. G. Sankar, and J. P. Liu, Adv. Mater. **23**, 821 (2011).

[13] N. T. Trung, Z. Q. Ou, T. J. Gortenmulder, O. Tegus, K. H. J. Buschow, and E. Brück, Appl. Phys. Lett. **94**, 102513 (2009).

[14] D. T. C. Thanh, E. Brück, N. T. Trung, J. C. P. Klaasse, K. H. J. Buschow, Z. Q. Ou, O. Tegus, and L. Caron, J. Appl. Phys. **103**, 07B318 (2008).

[15] See supplemental material for X-ray diffraction patterns of $Mn_{1.30}Fe_{0.65}P_{0.50}Si_{0.50}$ and magnetic entropy change of $Mn_{1.95}P_{0.50}Si_{0.50}$.

[16] K. A. Gschneidner and V. K. Pecharsky, Int. J. Refrig. **31**, 945 (2008).

[17] L. Häggström, L. Severin, and Y. Andersson, Hyperfine Interactions **94**, 2075 (1994).

[18] E. K. Delczeg-Czirjak, L. Delczeg, M. P. J. Punkkinen, B. Johansson, O. Eriksson, and L. Vitos, Phys. Rev. B **82**, 085103 (2010).

[19] S. Nagase, H. Watanabe and T. Shinohara, J. Phys. Soc. Jpn. **34**, 908 (1973).

[20] X. B. Liu and Z. Altounian, J. Appl. Phys. **105**, 07A902 (2009).

**Supplemental Material**

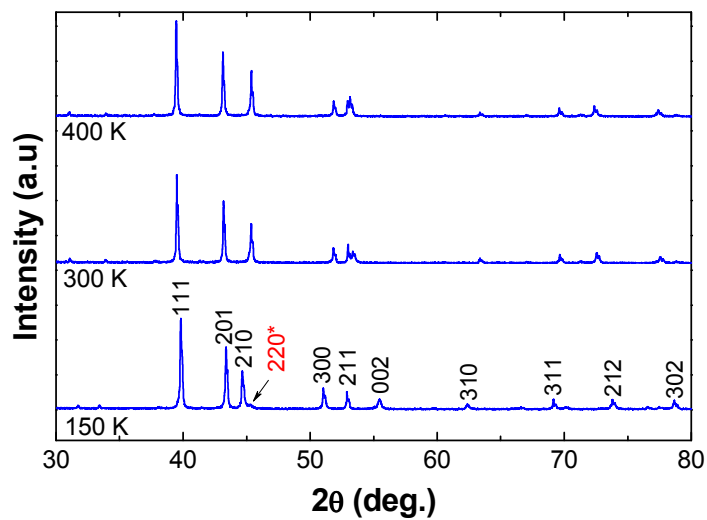

**Figure S1.** X-ray diffraction patterns measured in zero-field upon heating at 150, 300 and 400 K for $Mn_{1.30}Fe_{0.65}P_{0.50}Si_{0.50}$ compound. The patterns confirm that the sample crystallizes in the hexagonal $Fe_2P$-type structure. A very small amount of cubic $(Mn,Fe)_3Si$ impurity is also detected (*hkl* Miller index with asterisk (*) in red color). Above room temperature the peak of the impurity overlaps with that of the main phase.

**Supplemental Material**

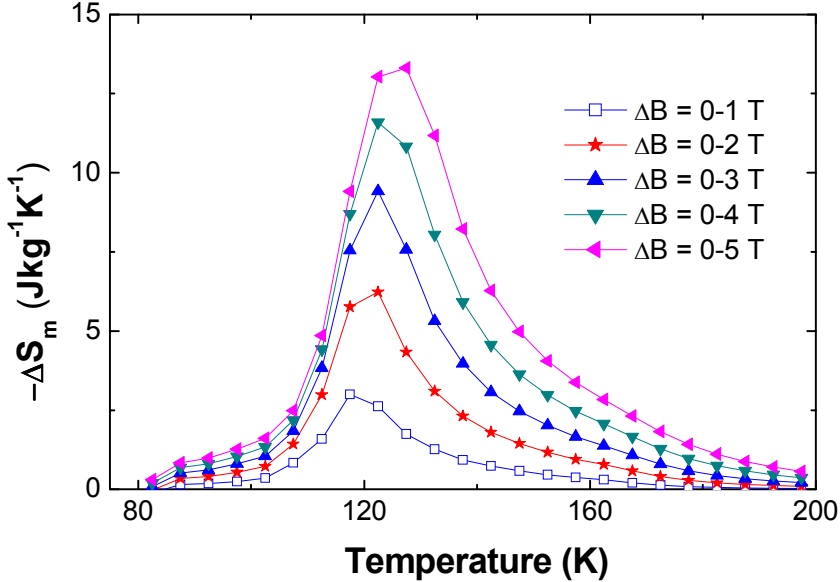

**Figure S2.** Magnetic entropy change as a function of temperature for $Mn_{1.95}P_{0.50}Si_{0.50}$.